\documentclass[a4paper,twocolumn,11pt]{quantumarticle}
\pdfoutput=1
\usepackage[utf8]{inputenc}
\usepackage[english]{babel}
\usepackage[T1]{fontenc}
\usepackage{amsmath}
\usepackage{hyperref}
\usepackage{amsfonts}

\usepackage{tikz}
\usepackage{lipsum}

\newtheorem{theorem}{Theorem}

\begin{document}

\title{Quantum Fourier transform computational accuracy analysis}

\author{Marina O. Lisnichenko}
\affiliation{Innopolis University, Innopolis, Russia}
\orcid{0000-0002-2701-6083}
\email{m.lisnichenko@innopolis.university}
\author{Oleg M. Kiselev}
\email{o.kiselev@innopolis.university}
\orcid{0000-0003-1504-7007}
\affiliation{Innopolis University, Innopolis, Russia}

\maketitle

\begin{abstract}
  The quantum Fourier transform (QFT) is a critical subroutine in many quantum algorithms such as quantum phase estimation and factoring. In this work, we present a rigorous accuracy analysis of the QFT that identifies three natural sources of accuracy degeneracy: (i) discretization accuracy inherited from classical sampling theory (the Nyquist–Shannon or Kotelnikov theorem), (ii) accuracy degeneracy due to limited resolution in eigenvalue (phase) estimation, and (iii) accuracy degeneracy resulting from finite quantum resources. We formalize these accuracy degradation sources by proving two theorems that relate the minimal amplitude and eigenvalue resolution to the number of qubits. In addition, we describe a gate-level implementation of the QFT and present simulation results on small-scale quantum systems that illustrate our theoretical findings. Our results clarify the interplay between classical signal discretization limits and quantum hardware limitations, and they provide guidelines for the resource requirements needed to achieve a desired precision.
\end{abstract}

\section{Introduction}

The quantum Fourier transform (QFT) is a cornerstone of many quantum algorithms \cite{shor1994,kitaev1995,HHL2009}. In its ideal form, the QFT is a unitary transformation that maps computational basis states to superpositions with well‐defined phase factors. However, practical implementations of the QFT must contend with accuracy of both the inherent limitations of discrete sampling (as dictated by the Nyquist–Shannon \cite{nyquist1928,shanon1949} or Kotelnikov theorem \cite{kotelnikov1933} and phase precision) and from the finite quantum resources available (e.g., a limited number of qubits).

Coppersmith \cite{coppersmith2002approximate} described the approximate version of the Fourier Transform as a quantum Fourier transform (QFT) and connected the final precision to an estimation of the number of operations required.
The other authors provide numerical examples of such algorithm (Zaman {\it{et al}} \cite{zaman2023step} as an example).

Here we sequentially examine a computational accuracy in the quantum Fourier transform algorithm and connect it to  the existence number of qubits. 

Our analysis focuses on three accuracy degeneration sources:
\begin{itemize}
\item Discretization error. The classical Fourier transform requires a sufficiently high sampling rate (by the Nyquist condition) to avoid aliasing. In the quantum case the number of qubits determines the number of frequency bins , and an insufficient leads to spectral leakage.

\item Eigenvalue (Phase) Resolution. In quantum phase estimation, the QFT is used to resolve eigenphases to a precision that scales as . We derive conditions relating the amplitudes of the eigenstates to this resolution.

\item Finite resource error. Errors arising from the inherent limitations of quantum gate implementations and the finite number of qubits lead to corrections that are, in many cases, bounded by constants or logarithmic factors in.
\end{itemize}

We now summarize the structure of the paper. Section  \ref{sec:methods} reviews the necessary background in classical Fourier analysis and quantum computation. Section \ref{sec:realisation} presents the rigorous accuracy analysis, including formal statements and proofs of two main theorems. Section \ref{sec:implementation} describes the gate-level implementation of the QFT and the simulation methodology. The section  presents simulation results that support our theoretical conclusions, and Section \ref{sec:summary} summarizes our findings.

\section{Methods}\label{sec:methods}

The quantum Fourier transform (QFT) was inspired from discrete Fourier transform (DFT).

Let the signal be as follows:

\begin{equation}\label{eq:signal}
    x(t)=\sum_{j=-\infty}^\infty a_j e^{i2\pi \nu_j t},
\end{equation}

Then, the DFT appears as follows:

\begin{equation} \label{eq:_dft}
    X_k = \frac{1}{\sqrt{N}}\sum_{n=0}^{N-1} x_n e^{-j\frac{2\pi k}{N} n},
\end{equation}

Here, $X_k$ represents amplitude of the $k$-th frequency component of the original signal $x_n$. The exponential term $e^{-j2\pi k n/N}$ is known as the phase, $n$ is a time index and $k$ the frequency index and the relation $k/N$ is a frequency of the signal.

The representation of the quantum Fourier transform is the same except for the following feature: $N = 2^n$ for some integer $n$ - the number of qubits. For now we introduce qubit as a unit vector having 2 elements. And both vectors 
$X$ and $x$ are also unit. The section \ref{sec:realisation} provides more insight about unitary nature. The $N = 2^n$ property follows from the following phenomenon.

Let a phase from \ref{eq:_dft} be represented as a binary number $\theta = b_1b_2.b_3b_4b_5...,$

where for $k \in [1,\infty)$ $b_k \in \{0, 1\}$. Then, since the part before the decimal does not matter ($e^{2\pi i k} = e^{2\pi i k-1} = \dots = e^{2\pi i \cdot 1})$, then phases of the form 

\begin{equation}\label{eq:singal_binary}
    \theta = 0.b_1b_2b_3..,
\end{equation}

only available as a choice. Then double phase $2\theta = b_1.b_2b_3\dots\equiv0.b_2b_3\dots,$ fourth degree of phase power $4\theta = b_1b_2.b_3\dots\equiv0.b_3\dots$, and so on.

Thus, we conclude that QPE inherits the properties of DFT including such limitations as

\begin{itemize}
    \item Kotelnikov frequency property
    \item aliasing
\end{itemize}

Below we describe both of the phenomena.

\subsection{Kotelnikov frequency property} \label{subsec:Kotelnokov_theorem}
The Kotelnikov frequency property refers to the fact that any band limited signal can be completely reconstructed from its samples taken at a rate equal to or greater than twice the highest frequency component of the signal. This principle is also known as the Nyquist–Shannon sampling theorem \cite{kotelnikov1933}. The Kotelnikov theorem states that if  $f_s$ is the sampling rate and $f_c$ is the maximum frequency present in the signal, then

\begin{equation}\label{eq:Kotelnikov}
    f_s = 2f_c.
\end{equation}

 If a signal has been sampled at or above the double rate, it can be perfectly reconstructed using $sinc$ interpolation:
$$ x(t) = \sum_{n=-\infty}^{\infty} x(nT) \operatorname{sinc}\left(\frac{t-nT}{T}\right) $$
Here, $x(nT)$ represents the sample values, $T$ is the time interval between samples ($T = 1/f_s$), and $\operatorname{sinc}(x)$ is defined as:
\begin{equation}
\operatorname{sinc}(x) =
\begin{cases} \nonumber
1 & \text{if } x = 0 \\
\frac{\sin(\pi x)}{\pi x} & \text{otherwise}
\end{cases}
\end{equation}

From this paragraph we take the point about number of sampling frequencies: should be equal or above half of a number of input signal points. Thus, the success of the Fourier depends on sufficient number of Fourier series. The next section grasps more on effect of low sampling rate input signal decomposition. 

\subsection{Spectral leakage}\label{subsec:spectral_leakage}

Current section discovers the phenomena of spectral leakage in terms of QFT requirement for the input signal.

Let $a_j e^{2i\pi \nu_j t}$ be the input signal. While $\omega_k\not=\nu_j$, define the period for the signal $T_{jk}=1/(\nu_j-\omega_k)$. Suppose that $T=n_{jk} T_{jk}+t_{jk}$, where  $n_{jk}\in\mathbb{N}$ and $0\le t_{jk}<T_{jk}$. Then,  the Fourier transform of the signal is the following:

\begin{widetext}
\begin{equation}\label{eq:spectral_leak}
\begin{split} 
\int_{0}^{T} a_j e^{i\nu_j t} e^{-i\omega_k t} dt =
a_j\int_{0}^{n_{jk}T_{jk}} e^{2i\pi(\nu_j-\omega_k) t}  dt+a_j\int_{n_{jk}T_{jk}}^{n_{jk}T_{jk}+t_j} e^{2i\pi(\nu_j-\omega_k) t} dt=\\ 
a_j\frac{e^{2i\pi(\nu_j - \omega_k)t}}{2i\pi(\nu_j - \omega_k)}\Biggr|_{t = 0}^{t_{jk}} 
\Rightarrow 
\left|\int_{0}^{T} a_j e^{2i\pi(\nu_j-\omega_k) t} dt\right|< |a_j| t_{jk}.
\end{split}  
\end{equation}
\end{widetext}

The formula shows that if the frequency of Fourier frequencies $w_k\not=\nu_j$, then the transformation performs with an accuracy, which depends of the rest of dividing $T/T_{jk}$. 

In case $\nu_j=\omega_k$ the $t_{jk}=T$ and the value of the integral:
\begin{equation*}
    \int_{0}^{T} a_j dt = a_j T.
\end{equation*}

The same analisys for the DFT
\begin{equation} \label{eq:dft}
    X_{jk} = \frac{a_j}{\sqrt{N}}\sum_{n=0}^{N-1} e^{i2\pi(\nu_j -k)\frac{n}{N}},
\end{equation}
Let's estimate the value $X_{jk}$. As $N\to\infty$, the sum can be interpreted as a Riemann sum for an integral over the unit circle:

\begin{equation*}
\begin{split} 
\lim_{N\to\infty}\frac{1}{N}\sum_{n=0}^{N-1} e^{i2\pi(\nu_j -k)\frac{n}{N}}=\\
\int_0^{1}e^{i2\pi(\nu_j -k)\zeta}d\zeta=\frac{e^{2i\pi(\nu_j -k)}-1}{2i\pi(\nu_j -k)}
\end{split} 
\end{equation*}
While $0<\nu_j<N$, then $\max_{j}\min_{k}|\nu_j-k|=1/2$, then:
$$
\max_{j}\min_{k} |X_{jk}|\sim |a_j|\frac{2}{\pi}\sqrt{N}.
$$
A sum of minimal values of the spectral leakages is an order of $\sum_{j}|a_j|$. Then,  the amplitude of frequency $\nu_j$ will be observed in the signal if :
$$
a_j\sqrt{N}\gg \sum_{j} |a_j|.
$$
This inequality states that the frequencies with small amplitudes are not observe in the signal. 

With addition to the Kotelnikov theorem for binary $N$-length signal theorem comes:

\begin{theorem} \label{theorem:min_ampl_above_alias}
    For QFT the minimal amplitude of the input signal frequency should satisfy

    \begin{equation} \nonumber
        \min(a_k) \geq \frac{\sum_{k=0}^{m}|a_k|}{2^{n/2}}
    \end{equation}
\end{theorem}

Proof: let input signal (\ref{eq:signal}) to have $m$ frequencies $\omega_k$ matching with $a_k$ amplitudes. As far as spectral leakage for the frequencies (\ref{eq:spectral_leak}), then each amplitude $a_k$ introduces correction proportional to $a_k$. For the whole set of frequencies the accuracy degeneracy $\epsilon = \sum_{k=-\infty}^{\infty}|a_k|$. Thus, for a number of Fourier discretion $M$ the minimal input signal $k-th$ signal's amplitude $\min(a_k) \geq \sum_{k=-\infty}^{\infty}|a_k|/M$. Due to the facts that $n$ two-element vectors stores $2^n$ frequencies and QFT operates with discrete values, $\min(a_k) \geq \sum_{k=0}^{m}|a_k|/2^n$.

The last two paragraphs described the natural limits of the QPE inherited from DFT. The next paragraph works on limitations from the side of quantum computation - multiple input signal Fourier transform and limited number of computation resource.

\subsection{Multiple input signal accuracy}\label{subsec:multiple_input_signal}

The quantum computations operate on unit vectors and unitary operators. Once again, next section discover this statement in a more complete manner.

If the signal has the form of unit vector $e^{i\lambda_i t}$, then the combination of such signals form unitary operator:

\begin{equation} \label{eq: multiple_phases_signal}
    U = e^{iAt} = \exp{\left(i
    \begin{bmatrix}
        \lambda_1 & \dots & 0 \\
        \vdots & \ddots & \vdots \\
        0 & \dots & \lambda_n
    \end{bmatrix}
    t\right)}
\end{equation}

Further section explains why this form is in the field of paper's interest.

As equation (\ref{eq:singal_binary}) states, the signal is of binary representation. Thus, vector of length $2^n$, where $n$ is an amount of two-element vectors describes any binary signal of this form. Therefore, QFT provides a univalent decomposition of any signal's (\ref{eq:singal_binary}) phase $\theta$ that satisfies the following form:

\begin{equation}\nonumber
    \theta = \frac{p_1}{2} + \frac{p_2}{4} + \dots + \frac{p_n}{2^n} = \sum_{t=1}^{n}\frac{p_t}{2^t},
\end{equation}

where $n$ is a number of two-element vector and $p_t \in \{1, \dots, 2^n-1\} \in \mathbb{Z}$. In case of insufficient input signal amplitude (\ref{eq:signal}) decomposition the accuracy degeneracy $\epsilon = O(\log_2 n)$

From the Kotelnokov's theorem \ref{subsec:Kotelnokov_theorem} and multiple input signal QFT property (\ref{subsec:multiple_input_signal}) theorem comes:

\begin{theorem}\label{theorem:min-max_eigenvalue}
    If the precision degeneracy for eigenvalue estimation $\epsilon \leq 1/2^n$, where $n$ is the number of two-element qubits, then the relation of the minimal and maximal eigenvalue of the operator follows: $\lambda_{min}/\lambda_{max} = 1/2^{n-1}$
\end{theorem}

Proof:

Number of qubits need for success eigenvalue estimation is $n = O(\log_2(1/\epsilon))$, where $\epsilon$ is the willing estimation precision. Then amplitude of the minimal eigenvalues should be $\lambda_{min} > 1 / \epsilon$ or $\lambda_{min} > 1 / 2^n$. At the same time, following the Kotelnikov's theorem, the number of qubits twice less than amount of digits eigenvalue contains but it is not less than $1/2^n$. It means that precision degeneracy is twice less also - $2/2^n$. From here the minimal difference between $\lambda_{max}$ and $\lambda_{min}$ is $\lambda_{min}/\lambda_{max} = 2/2^{n}$  or $\lambda_{min}/\lambda_{max} = 1/2^{n-1}$ 

\section{Realization}\label{sec:realisation}
Quantum computations emerged from quantum mechanics. Starting from Schrodinger equation \cite{schrodinger1926undulatory}:

\begin{equation}\label{eq:schrodinger}
    i\hbar \frac{\partial}{\partial t} |\psi(t)\rangle = \hat{H} |\psi(t)\rangle,
\end{equation}

quantum computation operates on input vector $\psi(t)$ with operator $\hat{H}$ to obtain the output vector $i\hbar \frac{\partial}{\partial t} |\psi(t)\rangle$. Notation $|\cdot\rangle$ - ket - stands for column vector, also $\langle\cdot|$ - bra - row-vector exists.

In a context of QFT the Schrodinger equation can be written as follows:

\begin{equation}\nonumber
    |X\rangle = U |x\rangle,
\end{equation}

where $|x\rangle$ is an input signal, $U$ -- unitary operator responsible for QFT, $|X\rangle$ -- QFT frequencies set.

A unit vector in a two-dimensional complex vector space is a qubit \cite{nielsen2001quantum}.
Number of qubits dictates a length of $|X\rangle$ as $2^n$.

Kronecker multiplication $\otimes$ allows to obtain $|v\rangle$ of $n$ qubits as follows:

\begin{equation}\nonumber
    |v\rangle = \otimes_{k = 1}^{n} |\cdot\rangle_k
\end{equation}

Let be vectors $\vec{a}$ and $\vec{b}$ be of size $2$:

$$\vec{a} = \begin{pmatrix} a_1 \\ a_2 \end{pmatrix}, \quad \vec{b} = \begin{pmatrix} b_1 \\ b_2  \end{pmatrix}.$$

Then, Kronecker product $\vec{c} = \vec{a} \otimes \vec{b}$ is the following:

$$\vec{c} = \begin{pmatrix}a_1 \cdot \vec{b}\\ a_2 \cdot \vec{b} \end{pmatrix} = \begin{pmatrix} a_1 b_1 \\ a_1 b_2 \\ a_2 b_1 \\ a_2 b_2 \end{pmatrix}.$$

Each qubit evolves with time as (\ref{eq:schrodinger}) under the effect of unitary operators. In terms of multiple qubits in the left-hand side the operator on the right-hand side performs as Kronecker product of all operators affecting on each qubit. Thus, the Schrodinger equation obtains the following form:

\begin{equation} \label{eq:schrodinger_Kronecker}
    \otimes_{k = 1}^{n} |X\rangle_k = \otimes_{k = 1}^{n} U_k \cdot \otimes_{k = 1}^{n} |x\rangle_k.
\end{equation}

In case of $t>1$ the operator changes as well as ket states. The overall unitary operator performs as a dot product of Kronecker product operators in each separate moment of time (or step):

\begin{equation} \label{eq:operator_product}
    U = \prod_{k = 1}^n \otimes_{m=0}^{l}U_k^m
\end{equation}

The quantum Fourier transform with such unitary operators (or gates \cite{barenco1995elementary}) as Hadamard (H), phase (P), identity (I):

\begin{equation} \label{eq: H_P_gates}
\begin{split}
H = \frac{1}{\sqrt{2}} \begin{pmatrix}
1 & 1 \\
1 & -1
\end{pmatrix}\text{, } \\
P = \begin{pmatrix}
1 & 0 \\
0 & e^{i\frac{\pi}{n}}
\end{pmatrix}\text{, } 
I = \begin{bmatrix}
1 & 0 \\
0 & 1
\end{bmatrix}\text{, }
\end{split}
\end{equation}

the swap (SW) gate replaces $p$-th and $q$-th qubit with each other:

\begin{widetext}
\begin{equation} \label{eq:swap}
\begin{split}
SW_{p, q} = \sum_{i_1, \dots, i_n}|i_1 \dots i_q \dots i_p \dots i_n\rangle \langle i_1 \dots i_p\dots i_q \dots i_n| = \\
\sum_{i_1, \dots, i_n}|i_1\rangle\langle i_1| \otimes \dots \otimes
 |i_q\rangle\langle i_p|\otimes \dots \otimes |i_p\rangle\langle i_q|\otimes\dots |i_n\rangle\langle i_n|.
\end{split}
\end{equation}
\end{widetext}

The control operators extends mentioned gates affecting by control state to controlled. The operator $U$ controlled by qubit $m$ to qubit $n$ is called $C-U_n^m$ and performs as follow:

\begin{equation}\label{eq:control_gate}
\begin{split}
    C-U_n^m = \otimes_{i=0}^{m-1}I \otimes |0\rangle\langle 0| \otimes_{i=m+1}^{k} I +  \\
    + \otimes_{i=0}^{m-1}I \otimes |1\rangle\langle 1| \otimes_{i=m+1}^{n-1} I \otimes U \otimes_{i=n}^{k}I, 
\end{split}
\end{equation}

In particular control-phase ($C-P)$ spawns a phase-kickback effect: transfer a phase shift from one qubit to another. Equation below describes this phenomenon for $2$-qubit control $|c\rangle$ and target $|t\rangle$ case:

\begin{equation}\label{eq:phase_kickback}
\begin{split}
    |ct\rangle \xrightarrow{U_\phi}|c\rangle |t \rangle e^{i\phi}, \text{ if } c=1,\\
    |ct\rangle \xrightarrow{U_\phi}|c\rangle |t \rangle, \text{ if } c=0.
\end{split}
\end{equation}

Another linear, bounded, self-adjoint, idempotent operators $Q1$ and $Q2$ on Hilbert space $H$ are projectors on the orthogonal subspaces $H_1$ and $H_2$ respectively. In a common way, projector $P_n$ affects on qubit $|\psi \rangle$ as follows:

\begin{equation}\nonumber
   P_n|\psi\rangle = (\alpha|0\rangle + \beta|1\rangle)\cdot |n\rangle = \alpha|n\rangle,
\end{equation}

where $n \in \{0, 1\}$ for physical quantum computers:

\begin{equation} \label{eq: projectors}
    P_0 = |0\rangle \langle 0| \text{ , }
    P_1 = |1\rangle\langle 1|.
\end{equation}

Utilizing described gates and definition the next paragraphs below explain the QFT step by step.

\subsection{Data initialization}

QPE exploits binary basis \cite{mahmud2022efficient} for the computations, in which each qubit represents a bit of information. Binary encoding allows to encode numbers from $0$ to $2^n-1$ using $n$ qubits.

To achieve this feature the first step is to translate each qubit fr0m default state $|0\rangle$ to superposition using $n$ Hadamard gates \ref{eq: H_P_gates}.

The evolution of the state is the following:

\begin{equation} \label{eq:step_1_H}
    \otimes_{k=0}^{n}H|0\rangle^{\otimes n} = \Big(\frac{1}{\sqrt{2}}\Big)^n\sum_{x=0}^{2^n-1}|x\rangle.
\end{equation}

The phase gate (\ref{eq: H_P_gates}) transforms the superposition into state of input signal $e^{2\pi i \theta}$, where theta is a phase (\ref{eq:singal_binary}). The result of this step is the following:

\begin{equation} \label{eq:step_2_state preparation}
\begin{split}
    \otimes_{k=1}^{n}P(0.b_1\dots.b_k) \Big(\frac{1}{\sqrt{2}}\Big)^n\sum_{x=0}^{2^n-1}|x\rangle = \\
    \Big(\frac{1}{2^{\frac{n}{2}}}\sum_{k=0}^{2^n-1} e^{2\pi i\theta k}|k\rangle\Big).
\end{split}
\end{equation}

This is the outcome of the state preparation step. The next step is a quantum Fourier series calculation.

\subsection{quantum Fourier series calculation}

This chapter talks in short about quantum Fourier series, the original paper provides full description \cite{coppersmith2002approximate}.

The QFT of a single qubit performs as a Hadamard gate, because

\begin{equation}\nonumber
    H|x_j\rangle = \Big(\frac{1}{\sqrt 2}\Big)(|0\rangle + e^{2\pi i x_j 2^{-1}}|1\rangle)
\end{equation}

In case of multiple qubits control rotations inherit rotation angle effect  from the previous qubits. Such that input state evolves into the following:

\begin{widetext}
\begin{equation}\nonumber
\begin{split}
    QFT(|x_1x_2\dots x_n\rangle) = \frac{1}{\sqrt{N}}\big(|0\rangle + e^{2\pi i [0.x_n]}|1\rangle\big) 
    \otimes
    \big(|0\rangle + e^{2\pi i [0.x_{n-1}x_n]}|1\rangle\big)
    \otimes \\
    \dots
    \otimes
    \big(|0\rangle + e^{2\pi i [0.x_1x_2\dots x_n]}|1\rangle\big)
\end{split}
\end{equation}
\end{widetext}


In case of inverse QFT (iQFT) all the steps perform the same but in the opposite order. In case of iQFT the measurement step allows to set the amplitude for the given vector of phases. Next section explains the measurement step.

\subsection{Measurement}

The goal of the measurement is to extract the binary input signals of form (\ref{eq:singal_binary}) encoded as phases (\ref{eq: multiple_phases_signal}) and interpret the results accordingly.

In overall, $n$ qubits provide $2^{n}$ binary numbers. The combination of the projectors (\ref{eq: projectors}) allows to measure all the numbers from $0$ to $2^{n-1}$ in binary form:

\begin{equation}\nonumber
    \Tilde{\theta} = \frac{\otimes_{k=1}^{n}P_{i=\{0,1\}} \cdot QFT(x)}{2^n}.
\end{equation}

Each of the final output signals in decimal format is the following:

\begin{equation}\nonumber
    \theta_k = \frac{\Tilde{\theta}_{10}^k} {n},
\end{equation}

where $\Tilde{\theta}_{10}^k$ is the decimal form of a $k$-th measured binary value.

Measurements finalizes the algorithm execution. The next section contains the results of the QFT simulation.

\section{Implementation} \label{sec:implementation}

This chapter provides experimental observations of theorems \ref{theorem:min_ampl_above_alias} and \ref{theorem:min-max_eigenvalue}.

The modeling included simulation of the $4$ qubit QFT which makes it possible to estimate the $2^4 = 16$ input signals.

The figure below represents the simulation result for a single input signal and combination of $16$-signals with the same amplitude:

\begin{figure}[!ht]
\centering
\begin{minipage}{.45\textwidth}
\centering
\includegraphics[width=\linewidth]{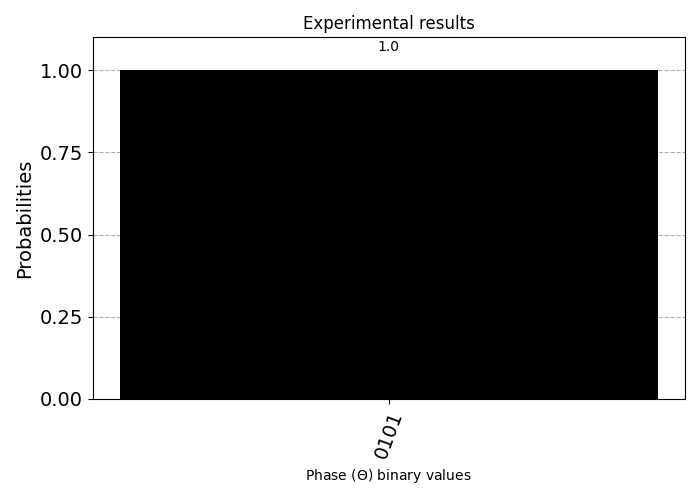}
\end{minipage}\hfill
\begin{minipage}{.45\textwidth}
\centering
\includegraphics[width=\linewidth]{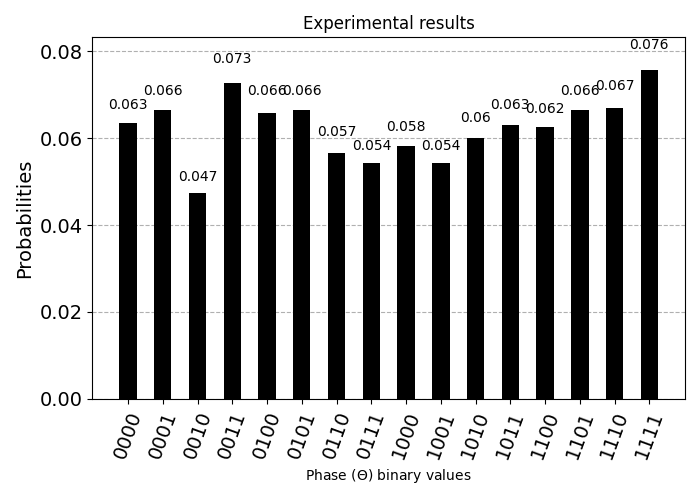}
\end{minipage}
\begin{otherlanguage}{english}
\caption{Simulation results for a single input signal (on the left)) and 16-channel signal (on the right). The x-axis shows the representation of the signal in a binary form and y-axis the measurement probability of that signal}
\label{fig:ideal_5_and_5_9}
\end{otherlanguage}
\end{figure}

The figure \ref{fig:ideal_5_and_5_9} shows that equal relation in signal amplitudes leads to equi-probability outcome. The next figure provides the result of algorithm execution for $3$ phases of values $\{3, 5, 7\}$ with the following relations of the amplitudes: $a_1/a_1 = 1/2$ and $a_1/a_3 = 1/4$:

\begin{figure}[!h]
\centering
\begin{minipage}{.45\textwidth}
\centering
\includegraphics[width=\linewidth]{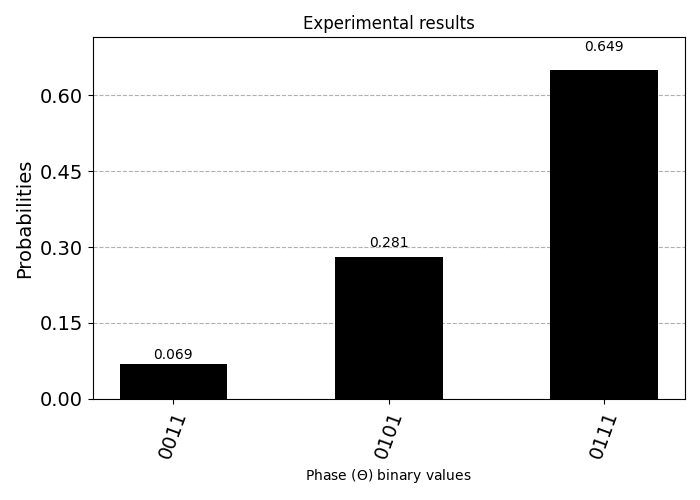}
\end{minipage}\hfill
\begin{minipage}{.45\textwidth}
\centering
\includegraphics[width=\linewidth]{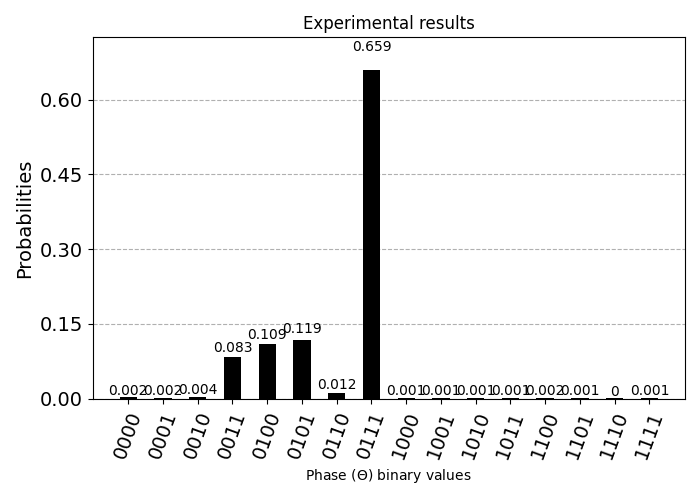}
\end{minipage}
\begin{otherlanguage}{english}
\caption{Simulation results for $3$ phases of values $\{3, 5, 7\}$ with the following relations of the amplitudes: $a_1/a_1 = 1/2$ and $a_1/a_3 = 1/4$ (on the left) and phases of values $\{2, 4.5, 7 \}$ with the same amplitudes (on the right). The non-integer phases spawn aliasing in the Fourier transform measurement}
\label{fig:3-5-7_alias}
\end{otherlanguage}
\end{figure}

The right side of the figure \ref{fig:3-5-7_alias} shows that relation of the probabilities are approximately equals to the squares relation of the input signal amplitudes if no frequency alias presented. The left side of the figure shows the histogram when number of Fourier frequencies (qubits) is not enough to describe the signal in definite manner. The experimental phase are $\theta = \{2, 4.5, 7 \}$ with the same amplitude relation. The results illustrate the theorem \ref{theorem:min_ampl_above_alias}: in case of $\theta=3$ the algorithm does not reflect the adequate information about this signal. And oppositely: in case of $\theta=7$ the high amplitude highlights this signal. 

Next histogram aims to prove the theorem \ref{theorem:min-max_eigenvalue} for the $\theta \in \{15, 17\}$.

\begin{otherlanguage}{english}
\begin{figure}[!h]
\centering
\begin{minipage}{.50\textwidth}
\centering
\includegraphics[width=\linewidth]{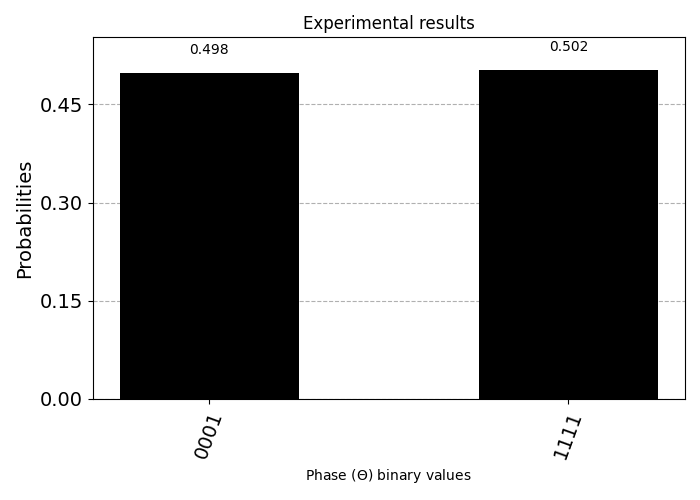}
\end{minipage}\hfill
\caption{Simulation results for phases of values$\{15, 17\}$. If the phase equals to $17$, the measurement gives $1$ as residue of $17/16 = 1_{10} = 0001_2$  }
\label{fig:15_17}
\end{figure}
\end{otherlanguage}

The figure \ref{fig:15_17} shows the result of the simulation. In case of $\theta=15$ the algorithm provides the expected result. However, for $\theta=17$ the algorithm provides the answer for the $2\pi + \theta$ what proves the theorem.

After all the algorithm description and experiments the next section draws the line and summarizes this study.

\section{Summary} \label{sec:summary}

We have presented an  accuracy analysis of the quantum Fourier transform, highlighting three key sources of accuracy degeneracy: discretization (aliasing) error, eigenvalue resolution error, and finite resource error. Two formulated theorems provide necessary conditions on the amplitudes of signal components and eigenvalue ratios in order to achieve reliable phase estimation with qubits.
Our simulation studies using Qiskit have shown the theoretical accuracy bounds and illustrated the sensitivity of the QFT to these accuracy degeneracy sources. 

\bibliographystyle{quantum}


\end{document}